\documentclass[acmsmall,pitch]{acmart}

\usepackage[T1]{fontenc}            
\usepackage[utf8]{inputenc}         
\usepackage{microtype}              
\usepackage[scaled=.8]{beramono}    
\usepackage[abbreviations]{foreign} 
\usepackage[frozencache,cachedir=.]{minted}
\usepackage[noabbrev,capitalise,nameinlink]{cleveref}
\usepackage[linesnumbered,ruled,vlined]{algorithm2e}
\usepackage{subcaption}
\usepackage{alltt}
\usepackage{url}
\usepackage{siunitx}
\usepackage{booktabs}
\usepackage{xcolor}
\usepackage{tikz}
\usetikzlibrary{shapes.geometric}
\usepackage[normalem]{ulem}

\definecolor{grey}{gray}{0.95}
\definecolor{ForestGreen}{RGB}{19,138,7}

\definecolor{mygray}{RGB}{240,240,240}
\usepackage[framemethod=TikZ]{mdframed}


\setcopyright{rightsretained}
\acmDOI{10.1145/3728966}
\acmYear{2025}
\acmJournal{PACMSE}
\acmVolume{2}
\acmNumber{ISSTA}
\acmArticle{ISSTA089}
\acmMonth{7}
\received{2025-02-27}
\received[accepted]{2025-03-31}

\begin{document}


\title{What Happened in This Pipeline? Diffing Build Logs with CiDiff}

\author{Nicolas Hubner}
\email{nicolas.hubner@labri.fr}
\orcid{0009-0005-4681-0353}
\affiliation{%
  \institution{Univ. Bordeaux, CNRS, Bordeaux INP, LaBRI, UMR 5800}
  \country{France}
}

\author{Jean-Rémy Falleri}
\orcid{0000-0002-8284-7218}
\email{falleri@labri.fr}
\affiliation{%
  \institution{Univ. Bordeaux, CNRS, Bordeaux INP, LaBRI, UMR 5800}
  \country{France}
}

\author{Raluca Uricaru}
\email{raluca.uricaru@labri.fr}
\orcid{0000-0002-5730-6428}
\affiliation{%
  \institution{Univ. Bordeaux, CNRS, Bordeaux INP, LaBRI, UMR 5800}
  \country{France}
}

\author{Thomas Degueule}
\email{thomas.degueule@labri.fr}
\orcid{0000-0002-5961-7940}
\affiliation{%
  \institution{Univ. Bordeaux, CNRS, Bordeaux INP, LaBRI, UMR 5800}
  \country{France}
}

\author{Thomas Durieux}
\email{thomas@durieux.me}
\orcid{0000-0002-1996-6134}
\affiliation{%
  \institution{TU Delft}
  \country{Netherlands}
}

\begin{abstract}
Continuous integration (CI) is widely used by developers to ensure the quality and reliability of their software projects.
However, diagnosing a CI regression is a tedious process that involves the manual analysis of lengthy build logs.
In this paper, we explore how textual differencing can support the debugging of CI regressions.
As off-the-shelf diff algorithms produce suboptimal results, in this work we introduce a new diff algorithm specifically tailored to build logs called CiDiff.
We evaluate CiDiff against several baselines on a novel dataset of \num{17906} CI regressions,  performing an accuracy study, a quantitative study and a user-study.
Notably, our algorithm reduces the number of lines to inspect by about \SI{60}{\percent} in the median case, with reasonable overhead compared to the state-of-practice LCS-diff.
Finally, our algorithm is preferred by the majority of participants in \SI{70}{\percent} of the regression cases, whereas LCS-diff is preferred in only \SI{5}{\percent} of the cases.
\end{abstract}

\maketitle

\section{Introduction}
\label{sec:introduction}

Continuous integration (CI) is widely used by developers to ensure the quality and reliability of their software projects~\cite{hilton_usage_2016}.
In a nutshell, CI involves compiling, testing, and applying a wide range of software quality tools with each code modification.
This process is organized into a series of steps known as a \emph{pipeline}.
These pipelines are executed by remote machines when using CI services, such as those offered by GitHub and GitLab.
The goal of a CI pipeline is to ensure that the project has no quality issues (\ie it compiles, all tests are passing, \etc).
Therefore, the primary outcome of pipelines is a resulting state, mainly \emph{passing} or \emph{failing}, allowing developers to be notified whenever a pipeline fails.

However, the resulting state of the pipeline only indicates \emph{when} there is a problem but never \emph{why} there is a problem.
Hence, developers need additional information when a pipeline suddenly fails to investigate the failure and fix the software project.
Unfortunately, pipelines run on remote machines and it is usually not possible to directly interact with them, preventing developers from using classical debugging tools to diagnose the failure.
To mitigate this issue, CI services also record extensive logs, allowing developers to perform a \textit{post-mortem} analysis of pipeline executions.
These logs mainly consist of outputs obtained from compilation, testing, and quality assessment processes.
We refer to them as \emph{build logs} in the remainder of this paper.

Debugging a build regression (\ie a build that was passing and suddenly failed) therefore involves analyzing the failing build log and looking for errors or warnings that could explain the failure.
However, build logs are lengthy and their exhaustive analysis is tedious.
In addition, looking for the cause of a failing build involves additional challenges.
The vocabulary used to signal a problem is not consistent between different build tools (\eg one may use \emph{error}, the other \emph{fatal}).
Some warnings or errors (\eg a package manager issuing a vulnerability warning) do not always trigger a failure, which can be misleading.
Finally, it is not unusual for a cascade of errors to span a large portion of the log, making it difficult to find them all.

Studying the changes between a passing and a failing version has long been known as an effective debugging technique for source code~\cite{wen_exploring_2019,an_fonte_2023}.
We explain in~\Cref{sec:motivation} how a textual diff~\cite{myers_ond_1986} between a passing log and a failing log could facilitate the debugging of a CI build failure.
However, we show in~\Cref{sec:motivation} that classical textual diff algorithms are not suitable for diffing build logs, causing their output to be riddled with irrelevant information and hindering the diagnosis.

In this paper, we introduce in~\Cref{sec:approach} a new diff algorithm tailored to build logs that produces better output with minimal configuration.
Our algorithm is based on a new log line similarity metric to determine the possible mappings between the lines of the two logs, combined with a bioinformatics-inspired heuristic to select the most relevant mappings.
The algorithm has been implemented in a diff tool, called CiDiff,\footnote{\url{https://github.com/labri-progress/cidiff}} which is notably able to pinpoint updated and/or moved lines, something that classical diff tools fail to deliver.
CiDiff empowers developers to understand the changes between two build logs with less effort.

In~\Cref{sec:evaluation} we evaluate our algorithm using a large dataset of pipeline regressions extracted from open-source projects hosted on GitHub, and compare it with the baseline LCS-diff algorithm~\cite{myers_ond_1986} and with two additional approaches: keyword-based search and bigram differencing.
Notably, our large-scale quantitative evaluation shows that our heuristic reduces the size of the diff compared to the LCS-diff baseline, simplifying error identification while maintaining a reasonable median run-time overhead over LCS-diff.

In summary, the contributions of this paper are:
\begin{itemize}
    \item 
    A novel bioinformatics-inspired heuristic to diff build logs notably able to detect updated and moved lines;
    \item An open-source implementation of this algorithm with a GUI tailored to assist developers in understanding a failure in a build regression;
    \item A novel dataset of \num{17906} pipeline regressions;
    \item An accuracy study, a quantitative study, and a user study comparing our heuristic to baseline approaches.
\end{itemize}

\section{Motivation}
\label{sec:motivation}
Figure \ref{fig:pair-of-log} illustrates a simplified excerpt of a common CI scenario that prompts a swift response from developers (colors in the figure can be ignored).
In this scenario, the CI pipeline of a software project works correctly until a newly introduced change in the project causes a failure.
To diagnose the issue, CI platforms such as GitHub Actions provide developers with extensive logs of the pipeline execution, allowing them to inspect the output to identify the root cause of the failure.
These \emph{build logs} assemble the output of the various tools involved at different steps of the pipeline execution (package managers, compilers, quality assessment tools, \etc).

However, because failures can arise from multiple sources, diagnosing CI failures is not always straightforward.
The failure-inducing change may originate not only from modifications to the software but also from changes to the pipeline specification itself.
Prior empirical research has shown that CI failures are diverse and can be triggered at any step of the pipeline~\cite{rausch_empirical_2017,vassallo_tale_2017}:~compilation errors, test errors, VCS errors, erroneous build meta-data, dependency resolution, code analysis, deployment, \etc.
While some approaches have been proposed to diagnose errors in specific build tools and steps (\eg~\cite{vassallo_every_2020,zhang_buildsheriff_2022,lebeuf_understanding_2018}), these do not account for the diversity of failures observed in practice and are not broadly applicable to \emph{any} build log.
This highlights the need for a generic approach to CI failure identification that remains entirely agnostic of the CI provider and the underlying tools involved in the pipeline.

As noted by \citeauthor{vassallo_every_2020}, developers typically rely on the error messages produced in CI logs for failure diagnosis~\cite{vassallo_every_2020}.
However, build logs are often lengthy and difficult to navigate, interleaving outputs from heterogeneous tools with inconsistent formatting and vocabulary, which complicates the identification of error messages.
In this paper, our intuition is that developers can compare the most recent \emph{passing log} before the change (\Cref{fig:passing_log}) with the new \emph{failing log} (\Cref{fig:failing_log}) to conduct a post-mortem analysis of the issue, identify the root cause of the failure, and fix the build.
Our hypothesis is that \textbf{textual differences between the last passing log and the failing log can assist developers in identifying relevant error messages}.
Specifically, because new error messages appear as a side effect of the failure and should not be present in the passing log, we hypothesize that \textbf{the important messages explaining the failure are most likely located within the added lines in the failing log}.

In this section, we first review existing approaches to log parsing and anomaly detection that have shown great success for production logs and discuss their applicability to build logs.
Then, we review other approaches that could be considered for identifying error messages in the failing log:~keyword-based identification and classical textual differencing algorithms.
Finally, we outline an overview of our custom textual differencing algorithm for build logs:~CiDiff.

\subsection{Production Logs Parsing and Analysis}
\emph{Production logs} consist of semi-structured text generated at run time by logging statements in software source code.
The growing volume of production logs has prompted significant research in the literature to enable effective monitoring of software systems by parsing logs into streams of structured events used to detect, predict, and remediate anomalies~\cite{he_survey_2021}.
Log parsing involves distinguishing between static tokens (or constants) and dynamic tokens (or variables) to extract recurring templates from log files.
Static tokens are typically extracted from logging statements in source code, although the extraction rules must be regularly updated following system updates~\cite{zhu_tools_2019,zhang_system_2023}.
The extracted templates can then be fed into feature extraction and machine learning algorithms to detect deviations and anomalies in new logs~\cite{he_survey_2021}.
Other approaches involve building finite-state models from logs to compare and visualize behavioral differences in the underlying protocols~\cite{amar_using_2018}.

Compared to production logs, build logs have received far less attention~\cite{korzeniowski_landscape_2022}.
They also exhibit specific characteristics that distinguish them from production logs and complicate the use of traditional production log parsing and analysis techniques.
While production logs are controlled by developers, build logs are essentially produced by third-party tools included in a CI pipeline (package managers, compilers, quality assessment tools, \etc).
Their format and evolution are thus not controlled by developers, complicating the identification of static and dynamic tokens necessary for extracting structured data.
Additionally, log mining techniques heavily rely on feature extraction and machine learning algorithms that require sufficient training data to be effective.
Other approaches highlighting behavioral differences and deviations in the underlying protocols are valuable for production logs but less so for build logs, where developers mainly rely on error messages to diagnose issues and fix their builds.
This highlights the need for a textual approach to build log differencing that does not make assumptions about the shape and content of build logs and can produce meaningful results from minimal input data.

\subsection{Keyword-based Failure Identification}
Without appropriate tooling, identifying the cause of a failure involves manually searching the failing log for relevant clues and error messages.
However, simple keyword-based approaches to finding such messages are often inadequate, as illustrated in the failing log of \Cref{fig:failing_log}.
In this log, keywords signaling potential problems vary from one line to another (\emph{warning} line \num{2}, \emph{failure} line \num{6}, no distinctive keyword line \num{7}, and \emph{error} line \num{10}).
These variations arise because CI pipelines rely on numerous third-party tools, each with its own distinct vocabulary.
As a result, a keyword-based search may suffer from poor recall, as demonstrated later in~\Cref{sec:accuracy}.
Moreover, keyword-matching messages can span the entire log (from line \num{2} to the last line in our example), requiring the developer to skim over the entire log.
Finally, another significant shortcoming is that messages matching the keywords are sometimes unrelated to the failure.
For instance, the message warning about potential vulnerabilities shown in line \num{2} is not related to the build failure caused by a parsing error in the file \texttt{foo.java}.
These unrelated error messages may confuse developers, causing them to waste time determining whether they are connected to the failure.

\subsection{Textual Differencing with LCS-diff}

To identify textual differences between the last passing log and the failing log, a natural solution is to turn to textual differencing algorithms.
Textual diffs effectively highlight differences, sparing developers from scrutinizing thousands of lines of logs, as illustrated by \Cref{fig:default-diff-shortcomings}.
This figure depicts so-called edit-scripts (or patches) generated using the Myers algorithm as implemented in the \texttt{diff} command~\cite{myers_ond_1986}, based on the longest common subsequence (LCS) between two sequences of text lines.
A pair of lines can be part of the LCS only if they are identical (unchanged).
The algorithm then categorizes the lines in the LCS as \emph{unchanged}, while other lines are labeled as either \emph{deleted} (if the line is only present in the reference passing log) or \emph{added} (if the line is only present in the failing log).
Note that the LCS-based algorithm requires unchanged lines to maintain the same relative order in both logs.

However, the algorithm described above produces suboptimal results when applied to build logs.
Indeed, it suffers from several shortcomings that result in an unnecessarily high number of added and deleted lines, slowing down the identification of relevant new error messages located in lines \num{6}, \num{7}, and \num{10} in the failing log of \Cref{fig:failing_log}.
First, it does not account for identical lines appearing in a different order in both logs, as unchanged lines in the LCS must have the same relative order in both files.
Yet, the order of lines in build logs is sometimes non-deterministic, with the sequence of actions being subject to change.
This is the case, for instance, for the downloading of dependencies before compilation, as shown in \Cref{fig:default-diff-shortcomings}, lines \num{1} to \num{5} in both logs.
The second major issue arises from the algorithm's limitation to three states for a given line: \emph{added}, \emph{deleted}, or \emph{unchanged}.
However, it is common for two build logs to contain ``equivalent'' lines, whose content changes due to the inclusion of a non-deterministic value, as shown in~\Cref{fig:default-diff-shortcomings}, lines \num{7} and \num{8} in the passing log and lines \num{8} and \num{9} in the failing log, where the duration and memory used differ between the two builds.
These two issues result in the inclusion of spurious added and deleted lines in the edit script, hindering the analysis of the changes.
To our knowledge, no method has been specifically developed to address these issues.

\subsection{CiDiff}
In this paper, we introduce a novel diff algorithm that tackles the above-mentioned issues, making it easier to identify new error messages inside a failing log as illustrated in~\Cref{fig:expected-diff}.
We believe that textual log diffing is a useful complement to existing log anomaly detection techniques, which require substantial knowledge (dedicated log parsers and/or supervised learning processes) to extract information from logs~\cite{he_survey_2021,bao_statistical_2019,locke_logassist_2022,vassallo_every_2020,zhang_buildsheriff_2022}.
It is also well-suited for build logs differencing, in contrast to production logs differencing approaches that focus on visually pinpointing differences in the order of events rather than the textual differences between the logs~\cite{amar_using_2018, bao_statistical_2019} (see~\Cref{sec:rw} for a more detailed comparison).

Our vision is to develop an approach that provides useful results regardless of the CI pipeline, requiring almost zero configuration and no prior knowledge beyond a passing log. This makes it highly practical to use.
Our approach acts as a \emph{swiss army knife}, supporting the debugging of failures in scenarios where no other approach is available.

\begin{figure*}
    \begin{subfigure}[b]{.48\textwidth}
        \centering\footnotesize
        \begin{minted}{diff}
 1 Downloading: com.fasterxml.jackson:core:2.9.0
 2 Warning jackson-core has 2 vulnerabilities
-3 Downloading: com.github.gumtreediff:core:2.1.0
 4 Downloading: ch.qos.logback:logback-classic:1.1

 5 Downloading: org.scala-lang:scala-library:2.11
-6 Compiling package core: success
-7 Total time: 5.170 s
-8 Final Memory: 19M/176M


            \end{minted}
            \caption{Passing log}
            \label{fig:passing_log}
        \end{subfigure}
        \begin{subfigure}[b]{.48\textwidth}
            \centering\footnotesize
            \begin{minted}{diff}
 1 Downloading: com.fasterxml.jackson:core:2.9.0
 2 Warning jackson-core has 2 vulnerabilities
 
 3 Downloading: ch.qos.logback:logback-classic:1.1
+4 Downloading: com.github.gumtreediff:core:2.1.0
 5 Downloading: org.scala-lang:scala-library:2.11
+6 Compiling package core: failure
+7 core/foo.java: unable to parse
+8 Total time: 5.361 s
+9 Final Memory: 19M/179M
+10 Error compiling project
            \end{minted}
            \caption{Failing log}
            \label{fig:failing_log}
        \end{subfigure}
    \centering
    \Description{Output of a classical diff between two simplified logs. Added (resp. deleted) lines are displayed in green (resp. red). It displays a moved line as being removed from the reference log (line \num{3}) and added at a different position in the modified log (line \num{4}); and three ``equivalent'' lines as being removed from the reference log (lines \num{6}, \num{7}, and \num{8}) and added in the modified log (lines \num{6}, \num{8} and \num{9}).}
    \caption{Output of a classical diff between two simplified logs. Added (resp. deleted) lines are displayed in green (resp. red). It displays a moved line as being removed from the reference log (line \num{3}) and added at a different position in the modified log (line \num{4}); and three ``equivalent'' lines as being removed from the reference log (lines \num{6}, \num{7}, and \num{8}) and added in the modified log (lines \num{6}, \num{8} and \num{9}).}
    \label{fig:default-diff-shortcomings}
\end{figure*}

\begin{figure}
    \centering
    \begin{subfigure}[b]{.48\textwidth}
        \centering\footnotesize
        \begin{alltt}
 1 Downloading: com.fasterxml.jackson:core:2.9.0
 2 Warning jackson-core has 2 vulnerabilities
 {\color{violet}3 Downloading: com.github.gumtreediff:core:2.1.0}
 4 Downloading: ch.qos.logback:logback-classic:1.1
 
 5 Downloading: org.scala-lang:scala-library:2.11
 6 Compiling package core: {\color{orange}success}

 7 Total time: {\color{orange}5.170} s
 8 Final Memory: {\color{orange}19M/176M}

        \end{alltt}
        \caption{Passing log}
    \end{subfigure}
    \begin{subfigure}[b]{.48\textwidth}
        \centering\footnotesize
        \begin{alltt}
 1 Downloading: com.fasterxml.jackson:core:2.9.0
 2 Warning jackson-core has 2 vulnerabilities
 
 3 Downloading: ch.qos.logback:logback-classic:1.1
 {\color{violet}4 Downloading: com.github.gumtreediff:core:2.1.0}
 5 Downloading: org.scala-lang:scala-library:2.11
 6 Compiling package core: {\color{orange}failure}
{\color{ForestGreen}+7 core/foo.java: unable to parse}
 8 Total time: {\color{orange}5.361} s
 9 Final Memory: {\color{orange}19M/179M}
{\color{ForestGreen}+10 Error compiling project}
        \end{alltt}
        \caption{Failing log}
    \end{subfigure}
    \Description{Desired output of a diff algorithm for the build logs of~\Cref{fig:default-diff-shortcomings}. Added (resp. deleted) lines are displayed in green (resp. red) while updated (resp. moved) lines are displayed in orange (resp. purple). Note that only the variable part is highlighted in orange in the case of updated lines.}
    \caption{Desired output of a diff algorithm for the build logs of~\Cref{fig:default-diff-shortcomings}. Added (resp. deleted) lines are displayed in green (resp. red) while updated (resp. moved) lines are displayed in orange (resp. purple). Note that only the variable part is highlighted in orange in the case of updated lines.}
    \label{fig:expected-diff}
\end{figure}

\section{Approach}
\label{sec:approach}

To address the limitations of the classical diff algorithm discussed in the previous section, we propose a novel diff approach specifically tailored for build logs. This approach introduces the following key improvements:

\begin{itemize}
    \item It identifies \emph{updated} lines (lines with marginal content changes but unchanged positions) in addition to the usual \emph{deleted}, \emph{added}, and \emph{unchanged} lines.
    \item It detects both \emph{moved-unchanged} lines (lines with unchanged content but different positions) and \emph{moved-updated} lines (lines with only marginal changes but different positions).
\end{itemize} 

\textbf{Identifying updated lines} (those that are conceptually equivalent to lines in the reference log but have minor changes) requires a custom similarity metric adapted to the structure of build logs. Traditional string similarity metrics are not appropriate here, as logs often feature highly structured lines and non-deterministic values with no effect on the semantics. 

Therefore, we introduce \textbf{a novel log line similarity metric} that accounts for the structure of the lines, providing a more accurate comparison than character-based metrics. Detailed explanations of this metric can be found in~\Cref{sec:sim-metric}.

For \textbf{detecting moved lines}, our new diff strategy incorporates a smart heuristic specifically designed to detect changes in the order of lines between a reference and a modified log, enabling accurate identification of moved lines. Since determining the optimal edit-script between two texts or between two line sequences (\textit{i.e.}, the smallest set of operations to transform one sequence into another) has been proved to be NP-complete if moves are allowed~\cite{shapira_edit_2007}, heuristics are crucial for efficient computation. 

To this end, we adapt techniques inspired by the genome alignment field in bioinformatics, namely \textit{seed-and-extend} strategies~\cite{hohl_efficient_2002,delcher_alignment_1999}. As genomes undergo large-scale evolutionary processes (\textit{i.e.}, rearrangements as moves of genomic parts and insertions/deletions of genomic material) and given their sizes, aligning genomic sequences -- a task that is closely related to computing an edit-script between two sequences -- is highly challenging. Genome alignment methods that deal with moves inside genomic sequences, 
typically implement \textit{seed-and-extend} strategies. \textit{Seed-and-extend} strategies work by quickly identifying highly similar parts of sequences (\textit{seeds}), which are likely to form part of the correct alignment~\cite{brudno_glocal_2003,darling_mauve_2004}.
These seeds are then extended and used to anchor the alignment, reducing the number of spurious matches between regions.

In our context, as detailed in~\Cref{sec:seed-extend}, we designed \textbf{an original seed-and-extend method} for log diffing. Here, the seeds are composed of groups of consecutive high-confidence matched lines, but these groups do not necessarily appear in the same order in both logs. This allows us to incorporate context while minimizing false-positive line matches. 

\subsection{Log Line Similarity Metric}
\label{sec:sim-metric}


To introduce the notion of updated line when performing the mapping between pairs of lines, we need a similarity metric tailored to the log diff context. Traditional string similarity metrics are ill-suited for our purposes because they treat lines as mere sequences of characters. However, lines in logs are more than just sequences of characters: they are composed of segments, where the order of the segments is important and conveys specific meanings. Indeed, lines in a log are created by a log statement, a piece of code that specifies the format of the string and which parts of the string are replaced by which value, such as the one in~\Cref{fig:log-statement}. So each line in a log has fixed segments (parts of the line that do not change with each iteration of the pipeline) and variable segments (parts of the line that might change with each execution of the pipeline). The line similarity metric we introduce exploits log line patterns to effectively capture semantic similarity while ensuring robustness to variations between lines. It is built upon the same fundamental assumptions as the popular log parsing tool, Drain~\cite{he_drain_2017}. Specifically, we leverage the observation that certain tokens remain consistent across similar log lines while others vary. Additionally, we assume that structurally similar log lines contain the same number of tokens, which aligns with Drain's parsing approach that groups log messages based on their number of tokens. In the following we summarize the similarity metric computation in Algorithm~\ref{alg:logsim} and give a complete explanation. 

\begin{figure}
    \centering\footnotesize
    \begin{minted}{java}
               log(String.format("%s Build %s, took %.1f seconds", now(), result, duration))
    \end{minted}
    \Description{Example of a log statement}
    \caption{Example of a log statement}
    \label{fig:log-statement}
\end{figure}

The segments of a line, also referred to as \textit{tokens}, are determined by splitting the line based on space characters. The space character serves as an effective delimiter in our context because of its role as a separator for natural language words. Log lines are typically written in natural language for human readability, and using space as a delimiter is consistent with this linguistic structure. Below we refer to fixed tokens as \textit{static tokens}, while variable ones are named \textit{dynamic tokens}. In addition, static and dynamic tokens will always have the same position in a line, regardless of the execution of the pipeline. For example in~\Cref{fig:log-statement}, the log statement has the following list of tokens: \texttt{"\%s"} (dynamic), \texttt{"Build"} (static), \texttt{"\%s,"} (dynamic), \texttt{"took"} (static), \texttt{"\%.1f"} (dynamic), and \texttt{"seconds"} (static).

When comparing pairs of lines we look at their organization into tokens and consider two aspects: their global structure (the concordance in the number of tokens) and their internal composition (the similarity between pairs of corresponding tokens). Note that pairs of identical lines are considered unchanged as in the classic diff algorithm, the following procedure applies only to pairs of lines displaying differences. Moreover, two lines with different numbers of tokens are considered different (\textit{i.e.}, similarity of \num{0}). 

We will further focus on pairs of lines that have the same number of tokens, since they may come from the same log statement. As two different log statements can produce the same number of tokens, we will also examine the similarity between their corresponding pairs of tokens: a token from the reference log line and the corresponding token at the same position in the line from the modified log. We impose that there is at least one pair of identical tokens (similarity of \num{1}), mostly corresponding to static tokens. Otherwise, the similarity of the lines is forced to $0$ as the two lines most likely do not come from the same log statement.  

The similarity between the remaining pairs of lines (the same number of tokens and at least a pair of identical tokens) is calculated as the average of the similarities between their token pairs. Pairs of identical tokens get a similarity of \num{1}.
To measure the similarity between two different tokens, two cases are examined; if one of these following two rules applies, the pair of tokens is considered \textit{similar} and gets a similarity of \num{0.5}, otherwise, the pair of tokens is considered \textit{different} and gets a similarity of \num{0}:

\begin{itemize}
\item First, equal length tokens are considered \textit{similar}. This is based on the observation that numerous dynamic tokens have a specific format and thus are represented on a constant number of characters. For example, a SHA-1 used by git always has \num{40} characters (2b185e5b87ad70f45b\-41c6346ddc0db1e33b8969), a timestamp is \num{19} characters long (2023-12-11T10:53:34), and an UUID is always represented on \num{32} characters (123e4567-e89b-12d3-a456-426614174000).

\item In a second time, tokens of different lengths are examined.
Some dynamic tokens, although different in length, do not vary much between two executions of the pipeline. For example, this is the case for the version number going from \texttt{1.2} to \texttt{1.2.1}. Here, two tokens are considered \textit{similar} if their tri-gram similarity is greater than or equal to a threshold $t^s$ . The tri-gram similarity between two tokens is computed based on the number of \num{3}-grams (substrings of length \num{3}) they have in common divided by the total number of \num{3}-grams present in the two tokens.
\end{itemize}

For example, the lines \texttt{2023-11-22 Build error, took 2.6 seconds} and \texttt{2023-11-23 Build success, took 3.4 seconds} have a token similarity of, respectively, \num{0.5}, \num{1}, \num{0}, \num{1}, \num{0.5}, \num{1}, which make the overall similarity of these lines \num{0.66}.

\begin{algorithm}[ht]
\caption{Logsim algorithm\label{alg:logsim}}
\DontPrintSemicolon
\KwData{S, T two strings representing two lines of log}
\KwResult{a real value in $[0,1]$}
\SetKwArray{U}{U}
\SetKwArray{V}{V}
\SetKwData{Tokens}{tokensSimilarity}
\SetKwData{Anchor}{hasStaticToken}
\SetKwIF{IfNot}{ElseIfNot}{}{if not}{then}{else if not}{}{}
\SetKwFunction{Split}{split}
    \DataSty{U} $\gets$ \DataSty{S}.\Split{\texttt{/\textbackslash s+/}}\;
    \DataSty{V} $\gets$ \DataSty{T}.\Split{\texttt{/\textbackslash s+/}}\;
    \If{$\left| U \right| \neq \left| V \right|$}{
        \Return{$0$\;}
    }
    \Anchor $\gets false$\;
    \Tokens $\gets 0$\;
    \For{$i=0$ \KwTo $\left| U \right|-1$}{
        \uIf{$U[i] = V[i]$}{
            $\Tokens \gets \Tokens + 1$\;
            $\Anchor \gets true$\;
        }
        \uElseIf{$\left| U[i] \right| = \left| V[i] \right|$}{
            $\Tokens \gets \Tokens + 0.5$\;
        }
        \ElseIf{$\left|\mathit{trigram}(U[i],V[i])\right| \geq t^s$}{
            $\Tokens \gets \Tokens + 0.5$\;
        }
    }
    \IfNot{$\Anchor$}{
        \Return{$0$\;}
    }
    \Return{$\frac{\mathit{tokensSimilarity}}{\left| U \right|}$\;}
\end{algorithm}

\subsection{Seed-and-Extend Algorithm}
\label{sec:seed-extend}

\begin{figure}[htbp]
    \centering
    \begin{subfigure}[b]{.45\textwidth}
    \centering\small
    \begin{alltt}
0  Fetching the repository
1  Remote: Counting object: 50\% (159/312)
2  Remote: Counting object: 100\% (312/312)
3  *[new ref] c1518c52f3
4  gradle build
5  Downloading datafixerupper-4.1.27.jar
6  Downloading guava-31.0.1.jar
7  Testing subproject 'core'
8  1 Test ok
9  Testing subproject 'common'
10 3 Tests ok
11 Tests SUCCESS
12 Build SUCCESS
13 duration: 8.66s
\end{alltt}
    \caption{Reference log}
\end{subfigure}
\begin{subfigure}[b]{.45\textwidth}
    \centering\small
    \begin{alltt}
0  Fetching the repository
1  Remote: Counting object: 100\% (318/318)
2  *[new ref] 531b38e562
3  gradle build
4  Downloading guava-31.0.1.jar
5  Downloading datafixerupper-4.1.28.jar
6  Testing subproject 'common'
7  Testing subproject 'core'
8  Warning: loose equality
9  4 Tests ok
10 Tests SUCCESS
11 Build SUCCESS
12 duration: 8.66s

\end{alltt}
    \caption{Modified log}
\end{subfigure}
    \Description{Pair of logs, on the left side the reference passing log and right side the modified failing log.}
    \caption{Pair of logs, on the left side the reference passing log and right side the modified failing log.}
    \label{fig:pair-of-log}
\end{figure}

\begin{figure}
    \centering
    \newcommand{\displaylog}[2]{
    \path node (A1) at (#1,7)   {$A$}   node (A2) at (#2,7)   {$A$}
          node (B1) at (#1,6.5) {$B_1$} node (B2) at (#2,6.5) {$B_2$}
          node (B3) at (#1,6)   {$B_3$} node (C2) at (#2,6)   {$C$}
          node (C1) at (#1,5.5) {$C$}   node (D2) at (#2,5.5) {$D$}
          node (D1) at (#1,5)   {$D$}   node (F2) at (#2,5)   {$F$}
          node (E1) at (#1,4.5) {$E$}
          node (F1) at (#1,4)   {$F$}   node (E2) at (#2,4.5) {$E$}
          node (G1) at (#1,3.5) {$G$}   node (J2) at (#2,4)   {$J$}
          node (H1) at (#1,3)   {$H$}   node (G2) at (#2,3.5) {$G$}
          node (J1) at (#1,2.5) {$J$}
          node (K1) at (#1,2)   {$K_1$} node (I1) at (#2,3)   {$I$}
          node (L1) at (#1,1.5) {$L$}   node (K2) at (#2,2.5) {$K_2$}
                                        node (L2) at (#2,2)   {$L$}
          node (M1) at (#1,1)   {$M$}   node (M2) at (#2,1.5) {$M$}
          node (N1) at (#1,0.5) {$N_1$} node (N2) at (#2,1)   {$N_2$}
    ;
}
\newcommand{\boxlink}[5][black]{	
	\node (X) [draw, #1, fill opacity=0, minimum width=1.2em, minimum height=#4em, label=left:#5] at (#2) {};
	\node (Y) [draw, #1, fill opacity=0, minimum width=1.2em, minimum height=#4em] at (#3) {};
	\draw[#1] (X.east) -- (Y.west);
}
\newcommand{\boxlinkb}[6]{
	\node (X) [fill opacity=0, minimum width=1.2em, minimum height=#5em, label=left:#6] at (#1) {};
	\node (a) [fill opacity=0, minimum width=1.2em, minimum height=1.2em] at (#3) {};
	\draw (X.south west) |- (X.north east);
	\draw (X.north east) -- (X.south east);
	\draw[blue] (a.south west) -| (X.south east);
	\draw[blue] (a.south west) -- (X.south west);

	\node (Y) [fill opacity=0, minimum width=1.2em, minimum height=#5em] at (#2) {};
	\node (b) [fill opacity=0, minimum width=1.2em, minimum height=1.2em] at (#4) {};
	\draw (Y.south west) |- (Y.north east);
	\draw (Y.north east) -- (Y.south east);
	\draw[blue] (b.south west) -| (Y.south east);
	\draw[blue] (b.south west) -- (Y.south west);

	\draw (X.east) -- (Y.west);
}
\newcommand{\boxlinkt}[6]{
	\node (X) [fill opacity=0, minimum width=1.2em, minimum height=#5em, label=left:#6] at (#1) {};
	\node (a) [fill opacity=0, minimum width=1.2em, minimum height=1.2em] at (#3) {};
	\draw (X.north west) |- (X.south east);
	\draw (X.south east) -- (X.north east);
	\draw[blue] (a.north west) -| (X.north east);
	\draw[blue] (a.north west) -- (X.north west);

	\node (Y) [fill opacity=0, minimum width=1.2em, minimum height=#5em] at (#2) {};
	\node (b) [fill opacity=0, minimum width=1.2em, minimum height=1.2em] at (#4) {};
	\draw (Y.north west) |- (Y.south east);
	\draw (Y.south east) -- (Y.north east);
	\draw[blue] (b.north west) -| (Y.north east);
	\draw[blue] (b.north west) -- (Y.north west);

	\draw (X.east) -- (Y.west);
}
\newcommand{\boxlinkbt}[8]{
	\node (X) [fill opacity=0, minimum width=1.2em, minimum height=#7em, label=left:#8] at (#1) {};
	\node (at) [fill opacity=0, minimum width=1.2em, minimum height=1.2em] at (#2) {};
	\node (ab) [fill opacity=0, minimum width=1.2em, minimum height=1.2em] at (#3) {};
    \draw (X.north east) -- (X.south east);
	\draw (X.north west) -- (X.south west);
    \draw[blue] (at.north east) -- (X.north east);
	\draw[blue] (at.north west) -- (X.north west);
    \draw[blue] (at.north east) -- (at.north west);
    \draw[blue] (X.south east) -- (ab.south east);
	\draw[blue] (X.south west) -- (ab.south west);
    \draw[blue] (ab.south east) -- (ab.south west);

	\node (Y) [fill opacity=0, minimum width=1.2em, minimum height=#7em] at (#4) {};
	\node (bt) [fill opacity=0, minimum width=1.2em, minimum height=1.2em] at (#5) {};
	\node (bb) [fill opacity=0, minimum width=1.2em, minimum height=1.2em] at (#6) {};
    \draw (Y.north east) -- (Y.south east);
	\draw (Y.north west) -- (Y.south west);
    \draw[blue] (bt.north east) -- (Y.north east);
	\draw[blue] (bt.north west) -- (Y.north west);
    \draw[blue] (bt.north east) -- (bt.north west);
    \draw[blue] (Y.south east) -- (bb.south east);
	\draw[blue] (Y.south west) -- (bb.south west);
    \draw[blue] (bb.south east) -- (bb.south west);

	\draw (X.east) -- (Y.west);
}
\begin{tikzpicture}[nodes={fill=white},align=center]
    \path node at (-0.75,7) {$[0]$}
		  node at (-0.75,6.5) {$[1]$}
          node at (-0.75,6) {$[2]$}
          node at (-0.75,5.5) {$[3]$}
          node at (-0.75,5) {$[4]$}
          node at (-0.75,4.5) {$[5]$}
          node at (-0.75,4) {$[6]$}
          node at (-0.75,3.5) {$[7]$}
          node at (-0.75,3) {$[8]$}
          node at (-0.75,2.5) {$[9]$}
          node at (-0.75,2) {$[10]$}
          node at (-0.75,1.5) {$[11]$}
          node at (-0.75,1) {$[12]$}
          node at (-0.75,0.5) {$[13]$}
    ;
    \displaylog{0}{1};  

    \path[draw,orange] (A1) -- (A2)
                (B1.east) -- (B2.west)
                (B3.east) -- (B2.west)
                (C1) -- (C2)
                (D1) -- (D2)
                (E1) -- (E2)
                (F1.east) -- (F2.west)
                (G1) -- (G2)
                (J1.east) -- (J2.west)
                (K1) -- (K2)
                (L1) -- (L2)
                (M1) -- (M2)
                (N1) -- (N2)
    ;
    \node (STEP0) at (0.5,-1) {All\\possible\\matches};
    \draw (STEP0.north west) -- (STEP0.north east);
    \draw[dotted] (1.5,5) -- (1.5,2);
    
    \displaylog{2.25}{3};

	\boxlink{2.25,7}{3,7}{1.2}{$t$}
	\boxlink{2.25,5.25}{3, 5.75}{2.5}{$u$}
	\boxlink{2.25,4.5}{3,4.5}{1.2}{$v$}
	\boxlink{2.25,3.5}{3,3.5}{1.2}{$w$}
	\boxlink{2.25,1.25}{3,1.75}{2.5}{$x$}

    \node (STEP1) at (2.75,-1) {1. Initial\\seeds\\selection};
    \draw (STEP1.north west) -- (STEP1.north east);
    \draw[dotted] (3.5,5) -- (3.5,2);

	\displaylog{4.25}{5};

	\boxlinkb{4.25,7}{4.97,7}{4.25,6.5}{4.97,6.5}{1.2}{$t$}
	\boxlinkt{4.25,5.25}{5.03, 5.75}{4.25,6}{5.03,6.5}{2.4}{$u$}
	\boxlink{4.25,4.5}{5,4.5}{1.2}{$v$}
	\boxlink{4.25,3.5}{5,3.5}{1.2}{$w$}
	\boxlinkbt{4.25,1.25}{4.25,2}{4.25,0.5}{5,1.75}{5,2.5}{5,1}{2.5}{$x$}

    \node (STEP2) at (4.75,-1) {2. Seed\\extension};
    \draw (STEP2.north west) -- (STEP2.north east);
    \draw[dotted] (5.5,5) -- (5.5,2);

    \displaylog{6.25}{7};

    \boxlink{6.25,7}{7,7}{1.2}{$t$}
	\boxlink{6.25,5.5}{7,6}{3.8}{$u$}
	\boxlink{6.25,4.5}{7,4.5}{1.2}{$v$}
	\boxlink{6.25,3.5}{7,3.5}{1.2}{$w$}
	\boxlink{6.25,1.25}{7,1.75}{5.4}{$x$}

    \node (STEP3) at (6.75,-1) {3. Seed\\overlap\\removal};
    \draw (STEP3.north west) -- (STEP3.north east);
    \draw[dotted] (7.5,5) -- (7.5,2);

	\displaylog{8.25}{9};

    \boxlink{8.25,7}{9,7}{1.2}{$t$}
	\boxlink{8.25,5.5}{9,6}{3.8}{$u$}
	\boxlink{8.25,4.5}{9,4.5}{1.2}{$v$}
	\boxlink{8.25,3.5}{9,3.5}{1.2}{$w$}
	\boxlink{8.25,1.25}{9,1.75}{5.4}{$x$}
	\boxlink[violet]{8.25,2.5}{9,4}{1.2}{$y$}
	\boxlink[violet]{8.25,4}{9,5}{1.2}{$z$}

    \node (STEP4) at (8.75,-1) {4. Additional\\seeds\\selection};
    \draw (STEP4.north west) -- (STEP4.north east);

\end{tikzpicture}
    \Description{Steps of the seed-and-extend algorithm applied on the pair of logs in \Cref{fig:pair-of-log}. Orange dashes denote every possible match between lines. Linked black squares depict the initial seeds and the blue parts correspond to the extension parts. The final seeds in black show seeds after overlap removal. The purple seeds are found during the selection of the additional seeds.}
    \caption{Steps of the seed-and-extend algorithm applied on the pair of logs in \Cref{fig:pair-of-log}. Orange dashes denote every possible match between lines. Linked black squares depict the initial seeds and the blue parts correspond to the extension parts. The final seeds in black show seeds after overlap removal. The purple seeds are found during the selection of the additional seeds.}
    \label{fig:seed-extends-steps}
\end{figure}
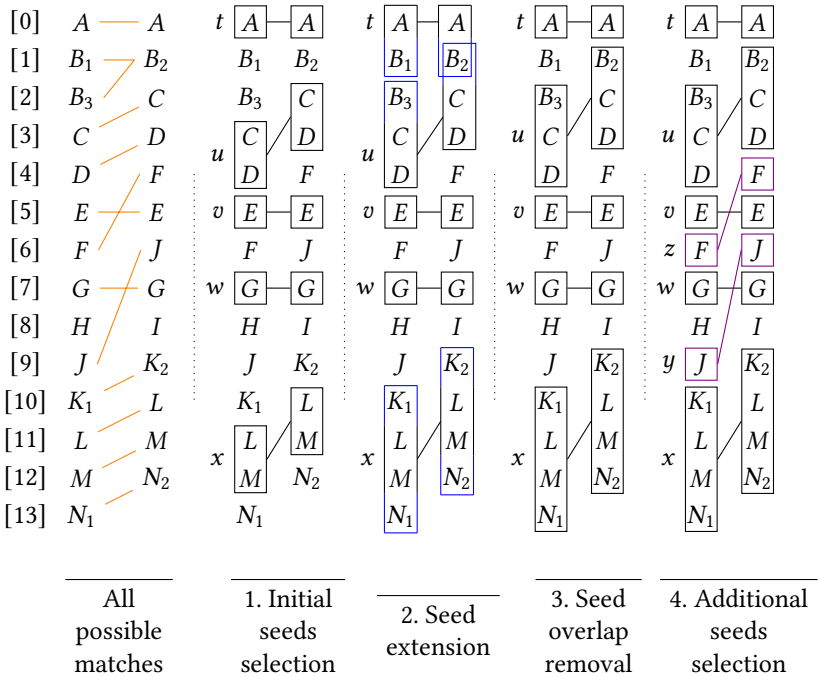

In this section, we introduce our new diff algorithm designed for build logs, which is based on a seed-and-extend heuristic.
Compared to the classical diff, our algorithm can detect identical and updated moved lines and produces an edit-script that considers these supplementary actions.
In total, our algorithm produces an edit-script with six actions: \textit{unchanged}, \textit{updated}, \textit{added}, \textit{deleted}, \textit{moved-unchanged} and \textit{moved-updated}.

The similarity between lines as computed by our similarity function described in the previous section, cannot produce a correct mapping of lines on its own because log lines are frequently duplicated, such as the test status lines depicted in~\Cref{fig:pair-of-log}.
To illustrate the complexity of the decisions one has to make regarding line matches based on line similarity alone, all possible matches between lines above a similarity threshold of \num{0.5} are depicted in~\Cref{fig:seed-extends-steps}.
To avoid spurious matches, contextual information (lines before and after a given log line) should be taken into account to make the best possible choices.
The seed-and-extend strategy we designed enables us to take into account this contextual information, as described below.

Our heuristic works in four main steps: (\num{1}) Initial seeds selection, (\num{2}) Seed extension, (\num{3}) Seed overlap removal and (\num{4}) Additional seeds selection. 

In the end, we convert the pairs of lines composing the final set of seeds into edit-script actions.
See below for full details on our seed-and-extend approach. 

For the remainder of the section, we introduce the following notations.
Couples of matching lines (with similarity above the threshold) are depicted by a pair $(X,Y)$ (where $X$ is an identifier), meaning that the line $X$ in the reference log is mapped to line $Y$ in the modified log.
Two identical lines are denoted by the same identifier, and similar lines (above a similarity threshold of $l^s$) by the same identifier but with a different subscript (like $(A_1,A_2)$).
After an extension step, a seed corresponds to a group of line pairs that are consecutive in both logs. It is represented by a triple $(r,m,s)$, where $r$ is the position of the first line of the block in the reference log, $m$ is the position of the first line in the modified log, and $s$ the size of the seed (the number of consecutive lines in each log).

For example, in~\Cref{fig:seed-extends-steps} (step \num{3}), the triple $(2,1,3)$ describes the seed containing the lines on positions $2,3,4$ from the reference log (\textit{i.e.}, $B_3$, $C$, $D$), which are paired respectively with lines on positions $1,2,3$ from the modified log (\textit{i.e.}, $B_2$, $C$, $D$).

\subsubsection{Initial seeds selection}
\label{sec:initialseeds}

The line pairs selected at the beginning of the procedure have a great influence on the pairings that follow in the matching process. This initial step is therefore crucial.
To avoid spurious matches that induce wrong pairings, we start by computing an LCS between the two logs with a classical diff algorithm, which gives us a set of high-confidence pairs of identical lines (for example, the lines named $E$ in~\Cref{fig:seed-extends-steps}). These selected pairs of lines appear in the same relative order in both log files. We further merge selected pairs of lines that are adjacent in both log files and construct our initial seeds. 

Step \num{1} in \Cref{fig:seed-extends-steps} depicts the $5$ seeds (\textit{i.e.}, $t, u, v, w, x$) computed at the end of this initial selection phase, for the logs of \Cref{fig:pair-of-log}.

\subsubsection{Seed extension}
\label{sec:seed-extension}
We continue the line-matching process by extending as much as possible the initial seeds computed in the previous step. Note that only pairs of lines meeting a similarity threshold $l^s$ and not already contained in an initial seed can be considered in the extension phase. For every initial seed, we apply the extension process in each direction by considering one by one the adjacent pairs of lines (both lines are either adjacent to the top of the seed in both logs or to its bottom). The extension of an initial seed will stop if it encounters an adjacent pair of lines with a similarity below the fixed threshold, or if it encounters a line that is part of another initial seed. 
Note that at the end of this step, unlike previously, extended seeds may overlap. However, such overlaps are restricted to lines having been added when performing the extension. 

For example, in step \num{2} of \Cref{fig:seed-extends-steps}, the initial seed $u$ is extended one line above (denoted by the blue rectangle). The extension of this seed stopped when reaching lines $B_1$ and $A$ as these lines do not match, and furthermore $A$ is already part of an initial seed.

\subsubsection{Seed overlap removal}
\label{sec:overlap-removal} 
As for the classical diff algorithm, the actions considered in our edit-script require a one-to-one mapping between the lines of both logs. However, during the seed extension phase, a line may be taken in two different seeds leading to overlaps between seeds. To ensure a one-to-one mapping between lines, in this phase, we apply a greedy strategy to remove such overlaps. Our overlap removal greedy strategy will prioritize large seeds (likely corresponding to high-quality, conserved blocks in the logs). The strategy consists of repeatedly selecting the largest seed and removing from the other seeds all line couples conflicting with a line contained in that seed. The process stops when there are no more overlaps between seeds. Above we observed that overlaps are due solely to lines added in the extension process. Therefore the overlap removal process cannot impact lines from the initial seeds. Moreover, by construction of the seeds, the overlap removal process cannot remove lines in the middle of the extended part of a seed.

Finally, newly adjacent seeds are merged. Two seeds $(r_1,m_1,s_1)$ and $(r_2,m_2,s_2)$ are adjacent if the following criterion is met: $(r_1+s_1=r_2 \land m_1+s_1=m_2) \lor (r_2+s_2=r_1 \land m_2+s_2=m_1)$. In other words, two seeds are merged if the end of one seed is adjacent to the beginning of the other. The merged seed is the union of the line couples of the two seeds.

The result of the overlap removal process is depicted in step \num{3} of \Cref{fig:seed-extends-steps}. First, the largest overlapping seed, $u$, is selected. Then the conflicting pair of lines $(B_1, B_2)$ is removed from the seed $t$ (as they are part of both $t$ and $u$ seeds).

\subsubsection{Additional seeds selection}
\label{sec:additionalseeds} 

In the end, we fill the gaps in-between the initial seeds by a applying an additional seed computation step that retrieves seeds that do not follow the order given by the LCS. Once excluding the lines taken in the initial seeds further extended and merged, remaining pairs of identical lines (the pairing is done based on the order in which they appear in both logs) having the same number of occurrences in both logs are now considered as seeds. These additional seeds undergo the same procedure as the initial ones, meaning seed extension with lines meeting the similarity threshold $l^s$, seed overlap removal and merging. This final step enables us to consider \textit{moved} lines (having changed position between the two logs), which allows us to introduce two supplementary actions in the edit script, namely \textit{move\_unchanged} and \textit{move\_updated}.

Such moved lines can be seen in the purple boxes in the step \num{4} of \Cref{fig:seed-extends-steps} (\textit{i.e.} the pairs of lines (J,J) and (F,F)).

\subsubsection{Computing the edit script}

Our seed-and-extend algorithm produces a set of seeds, but the edit script has yet to be generated. Given the one-to-one mapping property, seeds can be directly converted into actions. Every pair of lines in an initial seed (as at the end of the third step of the seed-and-extend strategy) becomes either \textit{unchanged} (in the case where the two lines are identical) or \textit{updated} actions. Note that as initial seeds are computed with an LCS algorithm, the included pairs of lines could not have undergone a move action. On the other hand, pairs of lines in the seeds computed during the additional seeds selection step are converted to \textit{moved-unchanged} or \textit{moved-updated} actions. The remaining lines (\textit{i.e.}, not inside a seed), become either \textit{deleted} if they come from the reference log or \textit{added} if they come from the modified log. 

\section{Evaluation}
\label{sec:evaluation}

In our evaluation we first assess the accuracy of our algorithm in comparison to the baseline LCS-diff and two additional approaches that were not specifically designed for textual diff but could be used in this context (bigram-based and keyword-based). This first evaluation was conducted on a sample of 100 manually annotated cases, randomly extracted from our complete dataset of 17,906 logs pairs, which correspond to pipeline regressions from real-world projects. 
We further focus on evaluating the output of our tool,  CiDiff, against the baseline LCS-diff on the complete dataset. Since no quantitative metric can serve as a perfect proxy for the quality of a diff algorithm's output~\cite{falleri_fine-grained_2024}, we complement this large-scale quantitative analysis with a user study, as is commonly done in diff algorithm evaluations~\cite{falleri_fine-grained_2014,decker_srcdiff_2020,falleri_fine-grained_2024,frick_generating_2018}.
Our evaluation protocol is detailed in the remainder of the section.
All code and data used in this evaluation are provided in our replication package.

\subsection{Dataset}
\label{sec:dataset}

To evaluate the efficiency of our log-diffing approach, we leverage the dataset of~\citeauthor{moriconi_improving_2024}~\cite{moriconi_improving_2024}, composed of consecutive raw CI logs. Compared to other existing CI log datasets that are usually small~\cite{brandt_logchunks_2020} or composed of processed logs~\cite{durieux_empirical_2020}, Moriconi's dataset contains approximately \num{500 000} GitHub Action raw logs from \num{28000} GitHub repositories that have at least \num{100} stars and at least \num{30} runs in the last \num{90} days (between the 1st and the 5th of October, 2023).

In our evaluation, we focus on the scenario where developers use log diffing to identify the cause of a regression. Therefore, we filter the dataset of Moriconi to extract only logs from pipelines where a regression appeared. More precisely, pairs of consecutive executions that changed state, from \textit{passing} to \textit{failing}, without any modification of the pipeline configuration. 


To find regressions inside Moriconi's dataset, we performed the following steps:
\begin{enumerate}
    \item Consecutive pipeline executions were paired;
    \item Pairs where the status of the first execution is passing and the status of the second execution is failing, while the workflow file remains unchanged, were extracted;
    \item Since pipelines might contain several job definitions, and since a pipeline execution fails as soon as one of the jobs it contains fails, we only retain the pairs of jobs whose executions went from passed to failed thus eliminating jobs that passed in both executions. 
\end{enumerate}

\begin{figure}
    \subfloat[Main programming language]{\label{fig:main_language_per_case}\includegraphics[width=0.48\linewidth]{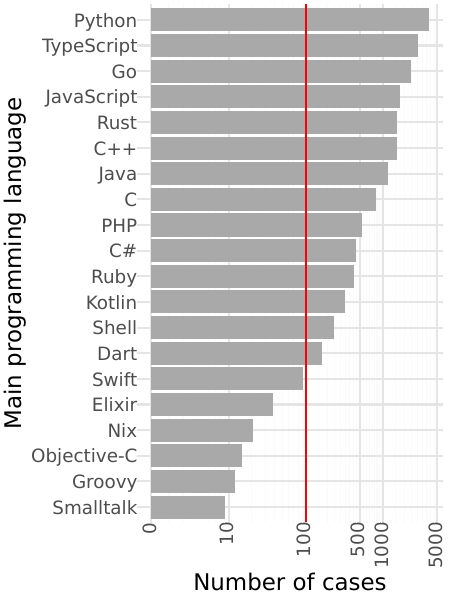}}%
    \hfill
    \subfloat[Total number of lines]{\label{fig:total_lines_per_case}\includegraphics[width=0.48\linewidth]{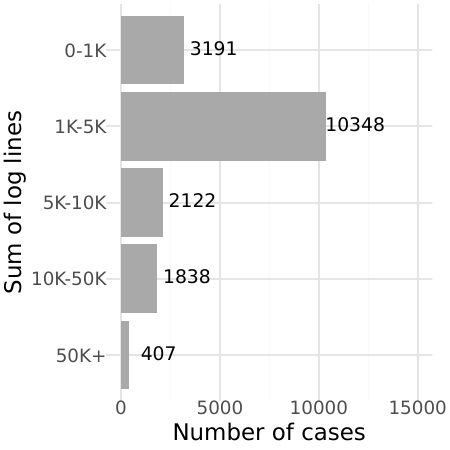}}\\
    \Description{Main programming language and total number of lines per case}
    \caption{Main programming language and total number of lines per case}
\end{figure}

In total, \num{17906} pairs of job-level build logs were identified, coming from \num{7955} repositories developed with one among \num{20} main languages as detected using GitHub's API.
The main language detected for each case is depicted in~\Cref{fig:main_language_per_case}.
We can see that \num{14} languages contribute more than \num{100} cases, with only Elixir, Groovy, Nix, Objective-C, Smalltalk, and Swift being below this limit.
Furthermore, the mean number of cases per repository is \num{2.25}, and the maximum number of cases contributed by a single repository is \num{60}.

Overall, our dataset includes cases from a very diverse range of projects and ecosystems.
\Cref{fig:total_lines_per_case} depicts the distribution of the sum of lines contained in the passing and failing logs for each case, with a median of \num{2183} (the first and third quartiles being \num{1213}, respectively \num{4877}).
However, some cases contain more than one million lines.

\subsection{Protocol}
\label{sec:protocol}

To evaluate CiDiff, we conducted three different studies. 
We first performed an \textit{accuracy study} to evaluate CiDiff's ability to pinpoint relevant error messages on a manually annotated subset of our dataset.
In this study, CiDiff is evaluated against three baselines: \num{1}) LCS-diff, which represents the state-of-practice in textual differencing, \num{2}) bigram differencing, the textual differencing approach at the core of \emph{s2KDiff} (one of the state-of-the-art methods for production log differencing~\cite{amar_using_2018}) and \num{3}) a keyword-based search where lines containing any of the following keywords are returned: \texttt{fail}, \texttt{error}, \texttt{exception}, \texttt{panic} illustrating the typical results expected from such a strategy.

We complement this accuracy study with an extensive \textit{quantitative study} on the entire dataset.
In this study, no ground-truth is available regarding the relevant error messages. Therefore, we use size-based metrics to evaluate the produced edit-scripts, with the underlying hypothesis that shorter diffs are easier to understand~\cite{martinez_hyperparameter_2023} than longer ones.
Size metrics alone are insufficient for evaluating the quality of a diff~\cite{falleri_fine-grained_2014,falleri_fine-grained_2024}, since a short edit-script can still include confusing actions. We therefore complement the quantitative study with a \textit{user study} involving nine participants, conducted on a subset of our dataset.
In the quantitative study and the user study, we use LCS-diff as the sole baseline, as it represents the state-of-practice in textual differencing.

Since our approach CiDiff has two hyper-parameters, we first proceed to a parameter tuning phase to explore and select optimal values for these hyper-parameters.
The selected values are used across the three studies.

\paragraph*{Parameter Tuning}
As explained in~\Cref{sec:approach}, our algorithm has two hyper-parameters: $l^s$ (line similarity threshold) and $t^s$ (token similarity threshold). Each threshold is a value between $0.0$ and $1.0$. Before running the evaluation, we inspected the variability of CiDiff's results with respect to the thresholds to determine the most suitable values for our algorithm.

We computed the precision and recall, as explained in the accuracy analysis paragraph (\ref{par:accuracy}) for each pair of $(l^s, t^s)$ ranging from $(0.0, 0.0)$ to $(1.0, 1.0)$ in increments of $0.1$ (\ie $(0.0, 0.0), (0.0, 0.1), ...,\\(1.0, 1.0)$). 
The results indicate that the highest precision is achieved when the line similarity threshold is between $0.1$ and $0.5$, and that the token similarity threshold has no significant effect within this range. However, outside of this line similarity threshold range, the token similarity begins to influence the results, where lower values lead to better precision.
Regarding the recall, every line similarity threshold yields very good results (with a median value between $0.7$ and $1.0$) and tends to improve as the threshold increases, except for $l^s=0.0$, where the recall is mostly below $0.5$. The token similarity has no impact on recall, except when set to $0.0$, which results in a slight decrease. The boxplots displaying the results for each configuration are available in the notebook of our replication package. 

Based on these results, we selected the configuration with $l^s=0.5$ and $t^s=0.6$ among those that yielded the highest precision.

\paragraph*{Accuracy Study}
\label{par:accuracy}

As detailed above, this experiment evaluates four algorithms: LCS-diff, unique bigram identification, keyword-based search, and our algorithm, CiDiff. 

We randomly sampled \num{100} cases in which one author manually annotated the failing log, marking all potentially relevant lines that could explain the failure.
To avoid excessively large logs, we only drew among cases where the log size was below \num{1000} lines.
We use recall and precision to assess how well the diff algorithms identify these relevant messages.
To compute the precision, we consider all the added lines as the output for the textual algorithms (LCS-diff, CiDiff, bigrams) and the lines containing the keywords for the keyword-based search.
Precision is then computed as the ratio of annotated lines present in the output lines over the total number of output lines. Recall is defined as the ratio of annotated lines present in the output lines over the total number of annotated lines.
We expect LCS-diff to have a recall close to \num{1} since it identifies all changes in the failing log as added lines. For the same reason, we expect LCS-diff to have a very low precision since most of the changes are probably not related to the failure.
Our expectation for CiDiff is to keep the recall close to \num{1} while increasing the precision (meaning that error messages were not placed in lines that were updated or moved).
In the results section we will analyze the distribution of the \num{100} recall and precision values across the \num{100} cases.

\paragraph*{Quantitative Study}

In the quantitative study, we will measure the size of the edit-script on each of the pairs of our complete dataset (described in Section~\ref{sec:dataset}) using the baseline LCS-diff and our heuristic CiDiff.
The size of the edit-script is the sum of the number of added, deleted, updated, moved-unchanged, and moved-updated lines in the two logs (the unchanged lines are omitted).
Note that for updated, moved-unchanged, or moved-updated lines, while they appear in both logs, they only count for 1 in the sum, favoring them over added and deleted lines.
For both algorithms, we pre-process the logs to remove the leading timestamps of each line.
In complement to the size of the edit-script, we will also measure the number of added lines in the failing log as we explained in~\Cref{sec:introduction} that these lines are essential to locate new error messages.
We hypothesize that smaller edit-scripts with fewer added lines are better to assist the debugging of regressions.
Since the ability to handle real-world logs is important, we will also measure the runtime of computing the diff between the passing and failing logs in milliseconds of both algorithms on all cases of our dataset.
The runtime is computed as the median value of three measures for each algorithm, when the first run is less than one minute (to account for the volatility of short runtime) or one measure if it is more than one minute.
To avoid exceedingly long running times, we set a \num{10} minutes timeout, meaning that if any of the measures exceed that duration, we stop measuring the runtime for that case and set the runtime as a dummy value (\num{-1.0}).

To report the results for the size of the edit-script and number of added actions while being able to compare them across the cases consisting of logs of different sizes, we will use the percentage difference defined as such: $p_m = 100 \times \frac{m_\mathrm{cidiff} - m_\mathrm{lcs}}{m_\mathrm{lcs} + 1}$ ($m$ being the metric of interest).
To avoid divisions by zero, \num{1} is added to the denominator, \num{1} being a relatively insignificant quantity compared to the size of an edit-script or the number of added lines.

A value $p_m$ of about \SI{-50}{\percent} indicates for instance that the value of CiDiff is half of the value obtained with LCS-diff, a value of about \SI{100}{\percent} indicates that the value of CiDiff is the double of the one obtained for LCS-diff.
To ensure that we have the same set of cases, we will use only the ones where both LCS-diff and CiDiff were not canceled due to a timeout.
Finally, we display the percentage differences across a set of bins computed according to the total log size (sum of the number of lines of the passing and failing logs).

To analyze the runtimes, we compare their distribution across the bins computed from the total log sizes to provide more information about the scalability of the algorithms.
We also discuss the number of timeouts for both algorithms.

\paragraph*{User Study}

In the user study, we conduct a human evaluation of CiDiff.
As manually assessing the diff produced on real build logs is highly time-consuming, we will restrict ourselves to analyzing \num{100} regression cases randomly drawn from the dataset, yielding a \SI{90}{\percent} confidence for a \SI{10}{\percent} error margin~\cite{thompson_sample_1987}.
Note that these \num{100} cases are not the same as the one used for the accuracy analysis since here we do not have to limit ourselves to logs smaller than \num{1000} lines.
In these cases, participants must identify the messages relevant to understanding the cause of the failure in the failing log based on the output produced by the two diff algorithms: LCS-diff versus CiDiff.

To provide the participants with a GUI supporting this scenario, we use the results of both algorithms in the following way.
We only show the failing log to the participants, augmented with colors on lines to indicate the changes compared to the passing log.
In addition to the classical green color to indicate added lines, we use the orange (resp. purple) colors to indicate an updated (resp. moved) line, as shown on the right side of~\Cref{fig:expected-diff}.
Note that there are no deleted lines in the failing log.
Build logs are typically very long, with only small portions being relevant to understanding the failure.
As explained in~\Cref{sec:introduction}, added lines are the most probable location for error messages indicating a failure.
Therefore, to filter as many irrelevant lines as possible in the view presented to the participants, we only display added lines (the green ones) along with a context of three lines above and below (that may correspond to any type of action).
All remaining unchanged, moved-updated, moved-unchanged and updated lines are hidden by default.
However, it is still possible to display these lines through a toggle in our GUI.
One such view is computed using LCS-diff and the other with CiDiff, but they are pseudonymized as alpha and beta.
Both views are displayed at the same time, with an alternating order to avoid a serial effect bias.
A short guide, provided to all participants, explains how to use this annotated log to search for relevant error messages (the error messages are most probably in the added lines of the log, and some error context might be in the added, updated or the moved lines).

The \num{100} cases are split into three sets of \num{33}, \num{33} and \num{34} cases.
As a judgment might be subjective, we present each case to three different participants, for a total of nine participants to the experiment (with no overlap between the three participants rating each set).
Participants were recruited in our lab but consisted only of persons not involved in this work.
A participant is asked for each case which view they would prefer in the scenario where they would need to find messages explaining the cause of the failure in this failing log.
The participants can choose alpha (LCS-diff), beta (CiDiff), or none if they believe neither view is better.
An aggregated judgment is then computed: LCS-diff or CiDiff if LCS-diff or CiDiff receives two or more votes out of the three, respectively; otherwise, the result is none.
We recruited a total of \num{9} participants, \num{3} with a bachelor level, \num{3} with a master level, and \num{3} with a PhD level in computer science.
Their programming experience spans \num{2} to \num{30} years, and their familiarity with diff tools, CI processes, and build logs spans from "Unused" to "Frequently used".

\subsection{Accuracy Study}
\label{sec:accuracy}

\begin{figure}
\subfloat[Precisions]{\label{fig:precisions} \includegraphics[width=0.49\linewidth]{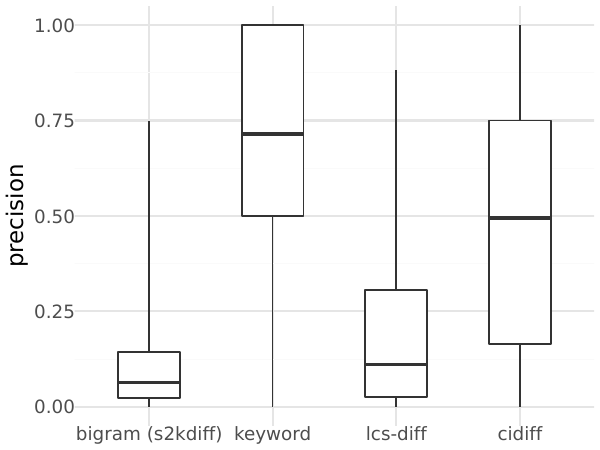}}
\hfill
\subfloat[Recalls]{\label{fig:recalls} \includegraphics[width=0.49\linewidth]{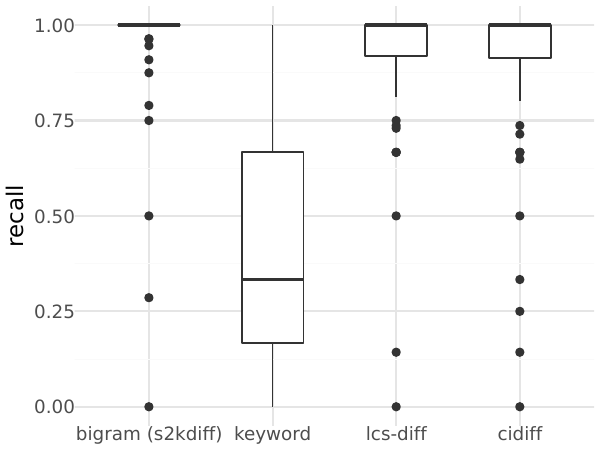}}
\Description{Boxplots of the recall and precision values of the four algorithms} 
\caption{Boxplots of the recall and precision values of the four algorithms} \label{fig:accuracy_results}
\end{figure} 

The precision and recall distributions of the bigram approach (the textual differencing at the core of s2Kdiff~\cite{amar_using_2018}), of the keyword-based search, of LCS-diff, as well as of CiDiff are shown in~\Cref{fig:accuracy_results}.
As conjectured, we can see that the recall of LCS-diff is very close to one with a mean of \num{0.93}.
Only in one case out of \num{100} LCS-diff was not able to identify any relevant log line.
Similarly, the recall of bigram is also very close to one (\num{0.97}).
As expected, the precision of LCS-diff is also very low with a mean of \num{0.14}.
The precision of bigram (\num{0.12}) is even lower than the one of LCS-diff.
CiDiff achieves a recall very close to the one of LCS-diff with a mean of \num{0.91}.
Similarly to LCS-diff, only in one case out of \num{100} CiDiff was not able to identify any relevant log line.
The precision of CiDiff is increased compared to LCS-diff and bigram, with a mean of \num{0.47}, which is more than double the precision of LCS-diff and bigram.
Finally, the keyword-based search exhibits an opposite trend: it has a good precision (mean of \num{0.68}) but a significantly lower recall (mean of \num{0.41}), illustrating the problems we described in~\Cref{sec:motivation}.
We argue that a low recall is a critical drawback in our use-case as failing to pinpoint a relevant log line ultimately leaves the user without assistance in locating it within the log.
In conclusion, these results demonstrate that CiDiff  effectively reduces irrelevant changes compared to LCS-diff and the bigram approach. It preserves relevant lines while filtering out a significant proportion of irrelevant ones.

\subsection{Quantitative Study}

\begin{figure}
    \subfloat[Size of the edit-script]{\label{fig:actions_diff} \includegraphics[width=0.49\linewidth]{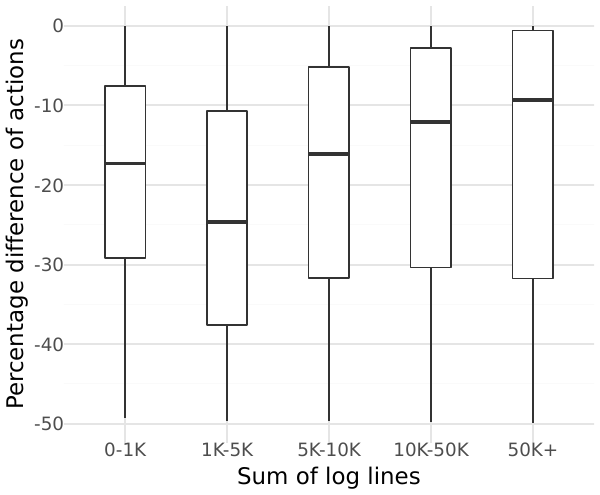}}%
    \hfill
    \subfloat[Number of added lines]{\label{fig:added_diff} \includegraphics[width=0.49\linewidth]{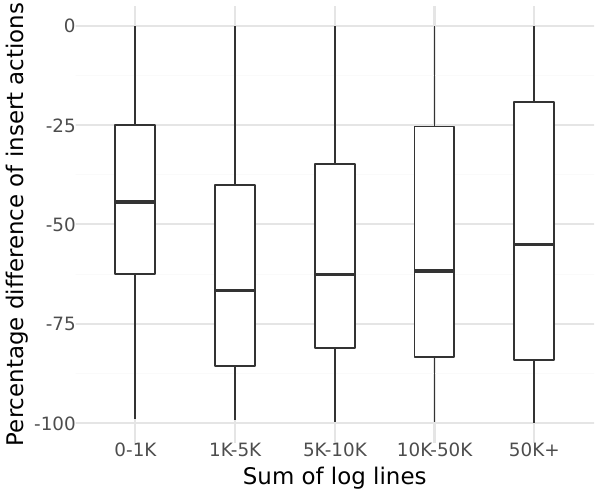}}%
    \hfill
    \Description{Percentage differences for the size of the edit-script and the number of added lines for CiDiff compared to LCS-diff}
    \caption{Percentage differences for the size of the edit-script and the number of added lines for CiDiff compared to LCS-diff}
\label{fig:results}
\end{figure} 

As depicted in~\Cref{fig:actions_diff}, the percentage differences in the size of the edit-scripts are between \SI{0}{\percent} and \SI{-50}{\percent}.
This means that CiDiff never produces a bigger edit-script than LCS-diff.
In the best cases, the size of the edit-script computed by CiDiff can be half the size of the one computed by LCS-diff.
In the median case, CiDiff computes a script \SI{20}{\percent} smaller compared to LCS-diff.
By analyzing the percentage differences across the log sizes, we observe that the reduction of the edit-script size is most significant for logs from \num{1000} to \num{5000} lines and decreases for larger logs.

If we zoom in on the number of added lines computed by both algorithms, we obtain the distribution of percentage differences depicted in~\Cref{fig:added_diff}.
This time, the percentage difference in the number of added lines is between \SI{0}{\percent} and \SI{-100}{\percent}.
Therefore, CiDiff never computes more added lines than LCS-diff and, in the best cases, can eliminate all added lines.
In the median case, CiDiff has an edit-script with about \SI{60}{\percent} fewer added lines than LCS-diff.
By analyzing the percentage differences across the log sizes, we note that the reduction of added lines is much more consistent than the reduction of the number of actions.
The best reduction is achieved for logs from \num{1000} to \num{5000} lines which are the most frequent in the dataset.

Overall, CiDiff consistently decreases the size of the edit-scripts and the number of added lines when compared to LCS-diff.

\begin{figure}
    \includegraphics[width=0.7\linewidth]{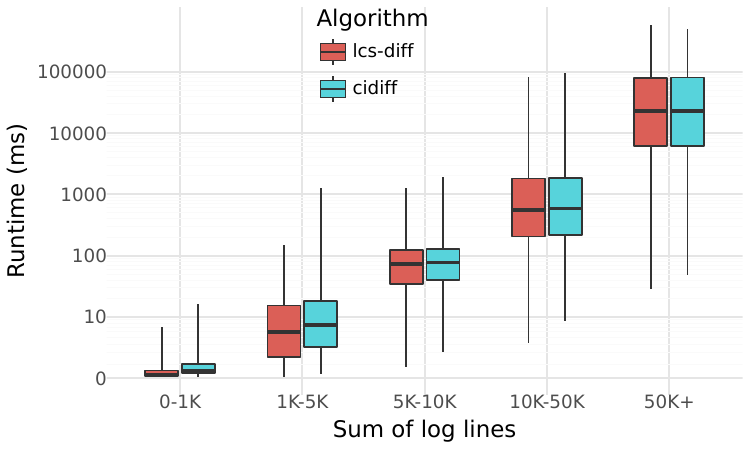}
    \Description{Distribution of runtimes of CiDiff and LCS-diff across the total lines of the cases}
    \Description{Distribution of runtimes of CiDiff and LCS-diff across the total lines of the cases}
    \caption{Distribution of runtimes of CiDiff and LCS-diff across the total lines of the cases}
    \label{fig:runtimes}
\end{figure}

Regarding the runtime, we experienced a timeout in \num{21} out of \num{17906} cases (\SI{0.11}{\percent}) both with CiDiff and LCS-diff.
The cases where timeouts occur are exactly the same for both algorithms: they are caused by a long running time for the computation of the LCS explaining why they occur for the same cases, since both algorithms rely on the computation of the LCS.
The distribution of the runtimes is shown in~\Cref{fig:runtimes}.
The first interesting result is that for all log sizes below \num{50000} the median runtimes of both algorithms are under one second.
We note a slight advantage for LCS-diff on small logs (between 0 and \num{5000} lines) with a slightly lower median runtime.
However, in these cases, both algorithms run very efficiently, with a median runtime below \SI{10}{ms}, making the processing time imperceptible to the user.
However, as log sizes increase, this advantage diminishes, and the runtimes for LCS-diff and CiDiff become very similar, indicating that the overhead of CiDiff compared to LCS-diff is negligible.
Finally, there are only five cases in the entire dataset where the runtime of CiDiff exceeds that of LCS-diff by more than a factor of \num{10}.
All of these cases involve logs ranging between \num{1000} and \num{10000} lines. 
A manual analysis of these cases revealed that they occur when CiDiff significantly reduces the number of actions (ranging from \SI{78}{\percent} to \SI{98}{\percent}).
However, the vast majority of cases with a similar reduction rate does not exhibit the same disparity in runtimes, suggesting that the structure of these five logs may represent a corner case for CiDiff.

\subsection{User Study}

As a result of our experiment with the participants, we received three ratings (out of \emph{CiDiff}, \emph{LCS-diff} or \emph{None}) per case, on \num{100} cases.
For each case, we aggregated the three ratings into a majority rating, as explained in~\Cref{sec:protocol}.
To assess the subjectivity of the experiment we computed a Fleiss Kappa measure of the inter-rater agreements on the \num{100} cases that yielded a \num{0.25} value (corresponding to a ``fair'' agreement).
This kappa value is rather low, confirming the need to use multiple raters and a majority rating.

\begin{figure}
    \includegraphics[width=0.5\linewidth]{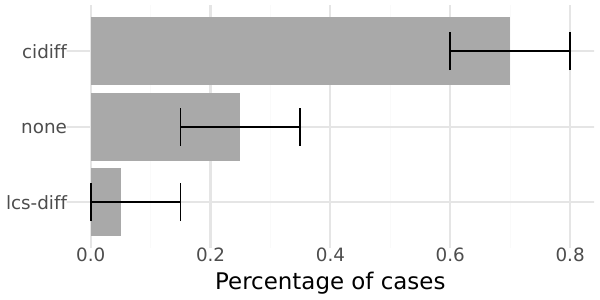}
    \Description{Favorite algorithm of the majority of users}
    \caption{Favorite algorithm of the majority of users}
    \label{fig:survey}
\end{figure}

The majority ratings proportion estimation on our random sample of \num{100} cases are displayed in~\Cref{fig:survey}, with the error bars corresponding to a \SI{90}{\percent} confidence interval.
We note that CiDiff was the favorite algorithm in \SI{70}{\percent} (confidence interval: \qtyrange{60}{80}{\percent}) of the cases.
In \SI{25}{\percent} (confidence interval: \qtyrange{15}{35}{\percent}) of the cases, no algorithm was preferred while in only \SI{5}{\percent} (confidence interval: \qtyrange{0}{15}{\percent}) of the cases LCS-diff was preferred.
This experiment confirms that shorter edit-scrips with fewer added actions as the ones of CiDiff are more useful for uncovering relevant error messages for understanding and debugging a failure.

\subsection{Threats to Validity}

Regarding the construct validity, as previously mentioned, the number of actions is not a perfect proxy for the quality of an edit-script.
For this reason, we ran a user study to compare the usefulness of the diffs.

Regarding internal validity, the manual annotation of relevant log lines for the accuracy analysis might bear some subjectivity.
Although the author who manually annotated the lines has significant expertise in CI systems, he is not an expert on the software systems involved in the 
dataset and might have made some mistakes.
Additionally, by having participated in the design of the algorithms, he might have favored annotations advantaging CiDiff.
To mitigate these biases, the author annotated the failing log independently, without looking at the passing log (therefore ensuring no prior knowledge of the results produced by LCS-diff or CiDiff).
The only assistance provided to the annotator was a GUI highlighting lines with an error keyword.
Similarly, the participants' judgment of the preferred algorithm in the user study is subjective and may be influenced by their lack of familiarity with the underlying systems.
Additionally, since the participants are from the same lab as the authors, they may have been biased in favor of CiDiff. To mitigate the impact of subjectivity, we involved three participants for each case and determined the preferred algorithm based on majority judgment.
Regarding the potential bias toward CiDiff, we could not fully anonymize the results, as CiDiff is easily distinguishable from LCS-diff due to its specific output format (including update and move actions).
However, to ensure transparency and allow for independent evaluation,  we provide the outputs of both LCS-diff and CiDiff in our replication package.

Regarding external validity, our dataset contains about \num{18000} cases from a very diverse set of projects and ecosystems hosted on GitHub.
However, we cannot claim that this dataset is representative of all CI technologies.
Additionally, our sample of \num{100} cases manually annotated in the accuracy study only includes logs with a total size below \num{1000} lines.
While this decision was made to keep the annotation process feasible, we acknowledge that our results may not fully generalize on longer logs.
Moreover, pipeline job logs on Github are generated by a single-thread, as is customary for pipeline runners, resulting in a relatively stable ordering of log messages.
However, our results may differ on distributed build systems, where log messages may appear in a more unpredictable order.
Finally, our human experiment involved only nine participants. While we aimed for a diverse range of experience levels among them, they do not represent a random sample of the broad population of CI users.

\section{Related Work}
\label{sec:rw}

\subsection{Continuous Integration and Build Logs}

Continuous Integration (CI) has become an essential practice in modern software development. 
\citeauthor{hilton_usage_2016} \cite{hilton_usage_2016} conducted a comprehensive study on the usage, costs, and benefits of CI in open-source projects. They found that CI helps projects release more often and is widely adopted by popular projects.
Build logs, which are generated during CI processes, have been recognized as a valuable source of information for developers.
\citeauthor{durieux_empirical_2020} \cite{durieux_empirical_2020} conducted an empirical study on restarted and flaky builds on Travis CI, highlighting the challenges developers face when dealing with CI failures and the importance of effective debugging techniques in this environment.

\subsection{Log Anomaly Detection}

\citeauthor{he_survey_2021} \cite{he_survey_2021} provided a detailed survey on automated log analysis for reliability engineering, covering the topic of log anomaly detection.
However, their focus was primarily on production logs rather than build logs.
A notable piece of information is that most approaches use log parsing and/or machine learning, which reduce their applicability to detect build anomaly in a CI context.
\citeauthor{korzeniowski_landscape_2022} \cite{korzeniowski_landscape_2022} conducted a systematic literature review and mapping study on automated log analysis, highlighting the limited research on build logs compared to production logs. This gap in the literature underscores the need for specialized tools and techniques for analyzing build logs.
Finally, some approaches~\cite{amar_using_2018,bao_statistical_2019} have been developed to diff production logs.
Their principle is to infer automata of log messages, to compute the differences between them and to visually highlight the differences on top of a combined automaton.
However, these approaches are designed to emphasize the differences in the order of messages, not the textual differences between the logs.
In a production log setting, such information is very valuable to quickly understand differences among the underlying protocols, but it is not adapted for build logs where developers search for textual error messages to enable debugging.

\subsection{Build Log Analysis}

Some researchers have explored techniques dedicated to build log analysis. \citeauthor{rosenberg_improving_2018} \cite{rosenberg_improving_2018} proposed an approach to improve problem identification via automated log clustering using dimensionality reduction. Their work demonstrated the feasibility of identifying problems in continuous deployment logs through clustering techniques.
Several approaches have been introduced to assist developers in identifying problems during builds~\cite{vassallo_every_2020,zhang_buildsheriff_2022,lebeuf_understanding_2018}.
However, these approaches are tailored to some specific build frameworks.
In contrast, our approach is applicable regardless of the build framework.
\citeauthor{brandt_logchunks_2020} \cite{brandt_logchunks_2020} introduced LogChunks, a dataset for build log analysis. This resource has been valuable for researchers working on build log-related problems, providing a common ground for experimentation and evaluation.
However, this dataset contains only \num{797} failing logs.
For our approach, we also need passing logs, preventing us from using this dataset.

\subsection{Diff Algorithms and Their Applications}

Diff algorithms have been widely used in software development for comparing different versions of files. \citeauthor{myers_ond_1986} \cite{myers_ond_1986} introduced the classic diff algorithm, which has been the foundation for many subsequent works in this area.
For source code, syntactic diff algorithms (i.e.~\cite{falleri_fine-grained_2014,fluri_change_2007,falleri_fine-grained_2024, decker_srcdiff_2020}) have been developed to provide more meaningful comparisons. Instead of relying on text-based algorithms, they use tree-based algorithms to take into account the syntactic structure of the code. However, these algorithms cannot be applied to build logs as they work on a tree structure that these logs do not exhibit.
Another type of approaches improve the classical text-based algorithm to identify moved or updated lines in the source code~\cite{canfora_ldiff:_2009, asaduzzaman_lhdiff:_2013}.
LDiff~\cite{canfora_ldiff:_2009} uses the LCS to select an initial set of unchanged lines.
Then it applies a two-step process where unmatched groups of lines are paired using a cosine similarity and for each pair of groups, lines are further paired using a Levenshtein similarity.
LHDiff~\cite{asaduzzaman_lhdiff:_2013} also uses the LCS to select an initial set of unchanged lines.
The remaining lines are matched in the cartesian product of all unmatched lines using a text similarity metric dedicated to source code lines (and similarity hashing to speed-up the computation).
In our work, we reuse the idea of using the LCS to select an initial set of unchanged lines. For the unmatched lines, we use a less expensive seed-and-extend strategy combined with a text similarity metric dedicated to log lines.
In the context of genomics, \citeauthor{hohl_efficient_2002} \cite{hohl_efficient_2002} developed algorithms for efficient multiple genome alignment, which share some similarities with the challenges faced in comparing build logs.

\section{Conclusion}

In this paper, we presented CiDiff, a novel textual diff algorithm dedicated to build logs as produced by CI pipelines.
Our primary objective for CiDiff is the scenario where developers have to debug a pipeline regression.
We examined the accuracy of CiDiff compared to three baselines and demonstrated CiDiff's good trade-off in terms of precision and recall. Moreover, we evaluated CiDiff's performance against LCS-diff on a novel dataset of \num{17906} CI regressions, using quantitative metrics.
Our algorithm reduces the edit-script size by about \SI{20}{\percent} and the number of added lines by about \SI{60}{\percent} in the median case while having a reasonable runtime overhead.
This reduction allows to better pinpoint error messages, with a significant precision improvement over LCS-diff while retaining a comparable recall.
Finally, on a random sample of \num{100} cases, our algorithm is favored by the participants of our user study in \SI{70}{\percent} of the cases, while LCS-diff is only favored in \SI{5}{\percent} of the cases.
As future work we plan to apply our approach to a broader range of use-cases than pipeline regressions.
A particularly interesting use case is matrix builds where a same pipeline is run across multiple environments.
We also plan to explore how parsers, including those from recent LLM-based approaches, could be leveraged to enhance our similarity metric. However, the computational overhead of such methods, as noted in~\cite{yu_brain_2023}, remains a challenge.

\section{Data Availability}

The code, data, and version of CiDiff used in our experiments are available in our replication package~\cite{zenodo_cidiff}. The latest versions of CiDiff can be found on its online repository (\url{https://github.com/labri-progress/cidiff}).

\begin{acks}
This work was partially funded by the French National Research Agency through grant ANR ALIEN (ANR–21–CE25–0007). We thank the anonymous reviewers for their precious comments.
\end{acks}

\bibliographystyle{ACM-Reference-Format}
\bibliography{biblio}

\end{document}